\documentclass[aps,groupaddress,nofootinbib,twocolumn]{revtex4}

\usepackage{graphicx}
\usepackage{amsmath}
\usepackage{epsfig}
\usepackage{hhline}
\usepackage{soul}
\usepackage{tabularx,pifont,multirow}
\usepackage{enumitem}
\usepackage{colordvi}
\makeatletter                                                         %
\newcommand*\rel@kern[1]{\kern#1\dimexpr\macc@kerna}                  %
\newcommand*\widebar[1]{                                              %
  \begingroup                                                         %
  \def\mathaccent##1##2{                                              %
    \rel@kern{0.8}                                                    %
    \overline{\rel@kern{-0.8}\macc@nucleus\rel@kern{0.2}}             %
    \rel@kern{-0.2}                                                   %
  }                                                                   %
  \macc@depth\@ne                                                     %
  \let\math@bgroup\@empty \let\math@egroup\macc@set@skewchar          %
  \mathsurround\z@ \frozen@everymath{\mathgroup\macc@group\relax}     %
  \macc@set@skewchar\relax                                            %
  \let\mathaccentV\macc@nested@a                                      %
  \macc@nested@a\relax111{#1}                                         %
  \endgroup                                                           %
}                                                                     %
\makeatother                                                          %

\include{Kaons_macro}

\begin{document}
\preprint[\leftline{KCL-PH-TH/2012-39, \, LCTS/2012-23, \,
 CERN-PH-TH/2012-243, MAN/HEP/2012/15}

%

\title{\Large {\bf Anomalous Majorana Neutrino
  Masses\\ from Torsionful Quantum Gravity } }

\bigskip

\author{\bf Nick E.~Mavromatos}

\affiliation{\vspace{1mm}
Theoretical Particle Physics and Cosmology Group,
  Department of Physics, King's College London, Strand, London WC2R
  2LS, UK and  Theory Division, Physics Department, CERN, Geneva 23,
  CH 1211, Switzerland} 

\author{\bf Apostolos Pilaftsis}

\affiliation{\vspace{1mm}
Consortium for Fundamental Physics,
School of Physics and Astronomy, 
University of Manchester,  Manchester M13 9PL, UK and             
Theory Division, Physics Department, CERN, Geneva 23, CH 1211,
Switzerland}


\begin{abstract}
\vspace{0.5cm}
\centerline{\bf Abstract }
\noindent\\[-2mm] 
The  effect of  quantum  torsion  in theories  of  quantum gravity  is
usually described by  an axion-like field which couples  to matter and
to gravitation  and radiation  gauge fields.  In  perturbation theory,
the couplings  of this torsion-descent  axion field are  of derivative
type and  so preserve  a shift symmetry.   This shift symmetry  may be
broken, if  the torsion-descent axion  field mixes with  other axions,
which could  be related to moduli fields  in string-inspired effective
theories.  In particular, the  shift symmetry may break explicitly via
non-perturbative  effects, when  these axions  couple to  fermions via
chirality  changing  Yukawa  couplings with  appropriately  suppressed
coefficients.  We~show, how in such theories an effective right-handed
Majorana neutrino mass can be  generated at two loops by gravitational
interactions that involve global anomalies related to quantum torsion.
We estimate the magnitude of the gravitationally induced Majorana mass
and find  that it is highly model dependent, ranging from multi-TeV to
keV scale.

\end{abstract}
\maketitle

\section{Introduction} 

The  recent discovery~\cite{:2012gk} of  the Higgs  boson at  the CERN
Large Hadron Collider (LHC) constitutes an important milestone for the
Ultra-Violet (UV) completion of the Standard Model (SM).  Although the
so-called Higgs mechanism  may well explain the generation  of most of
the particle masses in the SM, the origin of the small neutrino masses
still remains an open issue.  In particular, the observed smallness of
the  light neutrino  masses  may naturally  be  explained through  the
see-saw  mechanism~\cite{seesaw},   which  necessitates  the  Majorana
nature of the light (active)  neutrinos and postulates the presence of
heavy right-handed  Majorana partners of mass  $M_R$. The right-handed
Majorana mass $M_R$  is usually considered to be  much larger than the
lepton or  quark masses.  The  origin of $M_R$  has been the  topic of
several extensions of  the SM in the literature,  within the framework
of                            quantum                            field
theory~\cite{seesaw,Schechter:1980gr,Mohapatra:2005wg}    and   string
theory~\cite{Blumenhagen:2006xt}.

Recently, a potentially interesting radiative mechanism for generating
gauge-invariant  fermion  masses  at  three  loops  has  been  studied
in~\cite{ap}.   The mechanism utilizes  global anomalies  triggered by
the possible existence of scalar  or pseudoscalar fields in U(1) gauge
theories and  by heavy fermions $F$  whose masses may  not result from
spontaneous symmetry  breaking. One-loop quantum effects  of the heavy
fermions $F$ give rise to a global chiral anomaly given by
\begin{equation}
  \label{aFF}
a(x) \, F_{\mu\nu} (x) \, {}^\star\!F^{\mu\nu}(x) \; , 
\end{equation}
where $a(x)$  is a pseudoscalar  field, $F_{\mu\nu}$ denotes  the U(1)
gauge    field   Maxwell   tensor    and   ${}^\star\!     F^{\mu\nu}   =
\frac{1}{2}\,\varepsilon^{\mu\nu\rho\sigma}   F_{\rho\sigma}$  is  its
dual.  Moreover,  it was assumed in \cite{ap}  that the pseudo\-scalar
field   $a(x)$  couples   to  chirality-changing   fermion  bilinears,
$\widebar{f} i\gamma_5 f$, via the Yukawa-type couplings
\begin{equation}
  \label{yukawa}
y_a\, a(x)\, \widebar{f} i\gamma_5 f\; .
\end{equation}
The fermions $f$ are assumed to have zero bare masses. However, as was
explicitly demonstrated  in~\cite{ap}, the fermions $f$  can receive a
non-zero  mass   at  the  three-loop  level,   through  the  anomalous
interaction~(\ref{aFF})    and     the    chirality-changing    Yukawa
couplings~(\ref{yukawa}).

It  was  further  suggested  in \cite{ap}  that  this  mass-generating
mechanism can  also be applied  to create low-scale fermion  masses by
pure quantum  gravity effects.  In this  case, the r\^ole  of the U(1)
gauge field strength tensor~$F_{\mu\nu}$ will be played by the Riemann
curvature tensor~$R_{\mu\nu\rho\sigma}$,  and hence the  r\^ole of the
gauge fields by the gravitons.  Such a gravitationally-generating mass
mechanism could  straightforwardly be  applied to fermions  without SM
quantum  charges, such  as Majorana  right-handed neutrinos,  which we
restrict our attention here.  In  such a framework of quantum gravity,
the operator (\ref{aFF}) is expected  to be replaced by an operator of
the form
\begin{equation}
  \label{aRR}
a(x)\, R_{\mu\nu\rho\sigma} {}^\star\! R^{\mu\nu\rho\sigma}
\end{equation}
where  $a(x)$  is  an  appropriate  pseudoscalar  field  and  ${}^\star\!
R^{\mu\nu\rho\sigma}   =  \frac{1}{2}  \varepsilon^{\mu\nu\alpha\beta}
R_{\alpha\beta}^{\quad \rho\sigma}$ denotes the dual Riemann curvature
tensor.

It  is the  purpose of  this paper  to present  explicit  scenarios of
quantum gravity and provide  reliable estimates of the gravitationally
induced   right-handed   Majorana   mass  $M_R$.    Although   quantum
gravitational interactions  are non-renormalizable, nevertheless there
are aspects of the theory that can be exact, in a path integral sense,
and these are related to some aspects of torsionful manifolds. Torsion
appears as a non-propagating  form-valued pseudoscalar field $b(x)$ in
a quantum gravity  path integral and as such it  can be integrated out
exactly.  As  we explain  in this paper,  the effect of  torsion would
result  in  anomalous operators  analogous  to~(\ref{aRR}), which  are
instrumental in the generation of Majorana fermion masses.

Nevertheless, in addition to  the torsional field $b(x)$, the presence
of  extra pseudoscalar  fields $a(x)$  are required  for  generating a
chirality  violating Majorana  mass  $M_R$.  The  reason  is that  the
couplings of the  torsional axion field $b(x)$ are  of derivative type
and preserve  a shift symmetry:  $b(x)\to b(x) +  c$, where $c$  is an
arbitrary constant.   As a consequence, chirality is  conserved in the
massless  limit of  the  right-handed neutrinos,  thus forbidding  the
generation of a Majorana  mass~$M_R$.  However, the shift symmetry may
break  explicitly  via  non-perturbative  effects, when  these  axions
couple to  fermions via chirality  changing Yukawa couplings  $y_a$ of
the  form~(\ref{yukawa}).  The size  of the  Yukawa coupling  $y_a$ is
highly  model dependent,  implying  a  wide range  of  values for  the
gravitationally-induced Majorana mass scale~$M_R$.

This paper is organized  as follows.  After this introductory section,
Section~II  reviews some  basic properties  of manifolds  with quantum
torsion, within  field-theoretic and string-theoretic  frameworks.  In
Section~III we present  an explicit effective field-theoretic scenario
of  quantum  torsion that  can  give  rise  to an  anomalous  Majorana
neutrino mass generation at  two loops. Finally, Section IV summarizes
our conclusions and presents possible future directions.

\section{Properties of Quantum Torsion} 

Quantum  field  theories  in  space-times with  torsion  exhibit  some
interesting   properties,    which   have   been    known   for   some
time~\cite{torsion}.  In theories with fermions, torsion is introduced
necessarily in the first-order  Palatini formalism, where vierbein and
spin connections are treated as independent variables (Einstein-Cartan
theory)~\cite{kibble}.   In  what follows,  we  discuss various  cases
where torsion appears  in the spectrum of a  quantum-gravity model. We
shall examine first  a string theory models and  then proceed to argue
that certain properties  of quantum torsion, such as  those related to
chiral (axial) anomalies, are  generic to field theories with fermions
and  can arise  in ordinary  local  field theoretic  models, not  only
strings.  Let  us commence our  discussion from string theory,  not as
much  for  historical purposes,  but  because  the  latter provides  a
concrete UV complete theoretical framework for quantum gravity.

\subsection{Quantum Torsion and KR Axions in  String  Theories} 

In string theories, torsion is introduced as a consequence of the
existence of the antisymmetric tensor field $B_{\mu\nu} = -
B_{\nu\mu}$ existing in the gravitational multiplet of the
string. Indeed, as a result of the stringy ``gauge'' symmetry
$B_{\mu\nu} \rightarrow B_{\mu\nu} + \partial_{[\mu }B_{\nu]} $, the
low-energy string effective action depends only on the field strength
\begin{equation}
  \label{dB}
H_{\mu\nu\rho} = \partial_{[\mu} B_{\nu\rho]}
\end{equation}
where  the  symbol  $[\dots   ]$  denotes  antisymmetrization  of  the
appropriate indices. In fact, it can be shown~\cite{tseytlin} that the
terms involving the field strength perturbatively to each order in the
Regge slope parameter $\alpha^\prime$ can  be assembled, in such a way
that     only    torsionful     Christoffel    symbols,     such    as
$\widebar{\Gamma}^\mu_{\nu\rho}$,  appear.   In  this  formalism,  the
torsionful  Christoffel   symbol  $\widebar{\Gamma}^\mu_{\nu\rho}$  is
defined as
\begin{equation}
  \label{torsionful}
\widebar{\Gamma}^\mu_{\nu\rho}\ =\ \Gamma^\mu_{\nu\rho} +
\frac{\kappa}{\sqrt{3}}\, H^\mu_{\nu\rho}\ \ne\ 
\widebar{\Gamma}^\mu_{\rho\nu}\; ,  
\end{equation} 
where $\Gamma^\mu_{\nu\rho}  = \Gamma^\mu_{\rho\nu}$ is  the ordinary,
torsion-free, symmetric connection,  and $\kappa$ is the gravitational
constant given by
\begin{equation}
  \label{kappa}
\kappa^2 = 8\pi G_N = \frac{8\pi}{M^2_P}\ ,
\end{equation}
where  $G_N$ and  $M_P$ are  Newton's  constant and  the Planck  mass,
respectively.  Consequently,  terms involving the  generalized Riemann
curvature  tensor  ${\widebar  R}_{\mu\nu\rho\sigma}$, appear  in  the
effective    action.      In    the    context     of    (super)string
theories~\cite{strominger}, anomaly cancellation requires that Lorentz
and gauge  Chern Simons terms are  added to the field  strength of the
$\textbf{B}$  field  so that  one  may  defines  a new  field-strength
three-form $\textbf{H} =  \textbf{d B} + \frac{\alpha^\prime}{8\kappa}
\Big(\Omega_L  -  \Omega_V  \Big)$  such that  the  following  Bianchi
identity is implied:
\begin{equation}\label{bianchi} 
\textbf{d H } =\ \frac{\alpha '}{8 \kappa} {\rm Tr} \Big(\textbf{R}
\wedge \textbf{R} - \textbf{F}\wedge\textbf{F} \Big)\; , 
\end{equation}
where \textbf{R} denotes the gravitational Riemann curvature four-form
without H-torsion and $\textbf{F}$  the gauge field strength two form,
which includes  the torsion.  The $\wedge$  symbol denotes appropriate
contractions with the vierbeins $\textbf{e}^a_\mu$, while the trace is
taken over  all possible group-theoretic structures.   To lowest order
in $\alpha^\prime$,  where we shall  restrict our attention  here, the
effective  action  in a  four-dimensional  space-time (obtained  after
compactification and up to a total divergence) reads~\cite{tseytlin}:
\begin{eqnarray}
  \label{eat}
S^{(4)} &=& \int d^4 x \sqrt{-g} \Big( \frac{1}{2\kappa^2} R -
\frac{1}{6} H_{\mu\nu\rho}\, H^{\mu\nu\rho} \Big) \nonumber \\ 
&=&\int d^4 x  \sqrt{-g} \Big( \frac{1}{2\kappa^2} {\widebar R} -
\frac{1}{3} \kappa^2 H_{\mu\nu\rho}\, H^{\mu\nu\rho}  \Big)  \;,\quad
\end{eqnarray} 
where in the second line we used the generalized torsionful connection
(\ref{torsionful}). The gravitational constant $\kappa^2$ contains all
the  appropriate compactification volume  factors and  string coupling
terms, in particular we have the following relation from string theory
\begin{equation}
  \label{stringplanck}
\frac{1}{g_s^2} \, M_s^2 \, V^{(c)} = \frac{1}{2\kappa^2}\; , 
\end{equation}
with $M_s = 1/\sqrt{\alpha ^\prime}$ the string mass scale and $g_s$
the string coupling assumed weak $g_s < 1$. We assume constant
dilations $\phi $ in four dimensions for our purposes here. In
general, for non constant dilations $g_s = {\rm exp}(\phi)$ is a
field-dependent quantity.

In four dimensions, we may define the dual of $\textbf{H}$,
$\textbf{Y}~=~{}^\star\! \textbf{H}$, or equivalently in components,
\begin{equation}\label{baxion}
Y_\sigma\ =\ - 3\, \sqrt{2} \partial_\sigma b\ =\ \sqrt{-g}\,
\epsilon_{\mu\nu\rho\sigma} H^{\mu\nu\rho}\; , 
\end{equation}
after adopting the normalizations  of \cite{kaloper}. The field $b(x)$
is  a form-valued  pseudoscalar field,  with  a canonically-normalised
kinetic  term,  which  we  call  from  now  on  the  Kalb-Ramond  (KR)
axion~\cite{Kalb:1974yc},  in  order  to  distinguish  it  from  other
axion-like fields coming, \emph{e.g.}~from the moduli sector of string
theory,  which  we shall  also  employ  in  our analysis.   Using  the
definition  (\ref{baxion})   we  may  rewrite   the  Bianchi  identity
(\ref{bianchi}) in the form
\begin{equation}
  \label{Yder}
\nabla_\sigma Y^\sigma\ =\ \frac{\alpha ^\prime}{32 \kappa} \sqrt{-g}\,
  \epsilon_{\mu\nu\lambda\sigma} \Big( R_{a d}^{\,\,\,\,\,\,\mu\nu}
  R^{\lambda\sigma a d}  - F^{\mu\nu} F^{\lambda\sigma} \Big)\;,
\end{equation}
where  $\nabla_\sigma$  is  the torsion-free  gravitational  covariant
derivative  and $R_{\dots}$  are  the components  of the  torsion-free
curvature tensor.   The Latin indices $a,d$ are  tangent space indices
as usual.  Using (\ref{Yder})  in (\ref{eat}) and performing a partial
integration, we  arrive at the  following form of  the string-inspired
four-dimensional effective action with H-torsion~\cite{kaloper}:
\begin{eqnarray}
\label{string}
    S^{(4)} &=& \int d^4 x \sqrt{-g}\, \Big[ \frac{1}{2\kappa^2} R -
      \frac{1}{2} \partial_\mu b(x)\, \partial^\mu b(x) \nonumber \\ 
&&\hspace{-5mm}+\ \frac{\alpha ^\prime \,\sqrt{2}}{192 \kappa} b(x)
      \,\epsilon_{\mu\nu\rho\lambda } \Big(R_{a
        d}^{\,\,\,\,\,\,\mu\nu} R^{\rho\lambda a d } - F^{\mu\nu}
      F^{\rho\lambda} \Big)\Big]\; .\qquad
\end{eqnarray}
The close relation of the H-torsion to the appearance of an axion-like
field in the effective action is not unique to string theory.  In the
next subsection we proceed to discuss a field theoretical case where
similar effects take place.

\subsection{Quantum Torsion and KR Axions in Field Theory}

As  observed in  \cite{kaloper} in  the context  of QED  in torsionful
manifolds, one obtains similar  couplings of the torsion-induced axion
to  gravity  and  gauge  fields  as in  (\ref{string}),  by  employing
\emph{quantum anomalies} of the axial fermion current.  Indeed, let us
consider  Dirac QED  fermions in  a torsionful  space-time.  The Dirac
action reads:
\begin{equation}\label{dirac}
S_\psi = \frac{i}{2} \int d^4 x \sqrt{-g} \Big( \widebar{\psi}
\gamma^\mu \widebar{\mathcal{D}}_\mu \psi  
- (\widebar{\mathcal{D}}_\mu \widebar{\psi} ) \gamma^\mu \psi \Big)
\end{equation}
where $\widebar{\mathcal{D}}_\mu = \widebar{\nabla}_\mu - i e A_\mu $,
with  $e$ the  electron  charge  and $A_\mu$  the  photon field.   The
overline above  the covariant derivative, i.e.~$\widebar{\nabla}_\mu$,
denotes  the presence  of  torsion, which  is  introduced through  the
torsionful spin connection: $\widebar{\omega}_{a  b \mu} = \omega_{a b
  \mu} + K_{a  b \mu} $, where $K_{ab \mu}$  is the contorsion tensor.
The  latter  is  related  to  the  torsion  two-form  $\textbf{T}^a  =
\textbf{d   e}^a   +    \widebar{\omega}^a   \wedge   \textbf{e}^b   $
via~\cite{torsion,kaloper}:     $K_{abc}    =     \frac{1}{2}    \Big(
\textrm{T}_{cab}  - \textrm{T}_{abc}  -  \textrm{T}_{bcd} \Big)$.  The
presence  of torsion  in the  covariant derivative  in  the Dirac-like
action (\ref{dirac}) leads, apart from the standard terms in manifolds
without  torsion, to an  additional term  involving the  axial current
$J^\mu_5 \equiv \widebar{\psi} \gamma^\mu \gamma^5 \psi $:
\begin{equation}\label{torsionpsi}
S_\psi \ni  - \frac{3}{4} \int d^4 \sqrt{-g} \, S_\mu \widebar{\psi}
\gamma^\mu \gamma^5 \psi  = - \frac{3}{4} \int S \wedge {}^\star\! J^5  
\end{equation}
where $\textbf{S} = {}^\star\! \textbf{T}$  is the dual of \textbf{T}: $S_d
=   \frac{1}{3!}     \epsilon^{abc}_{\quad   d}   T_{abc}$.     

We  next remark that  the torsion  tensor can  be decomposed  into its
irreducible parts~\cite{torsion},  of which $S_d$  is the pseudoscalar
axial vector:
\begin{equation}
T_{\mu\nu\rho} = \frac{1}{3} \big(T_\nu
g_{\mu\rho} - T_\rho g_{\mu\nu} \big) - \frac{1}{3!}
\epsilon_{\mu\nu\rho\sigma} \, S^\sigma + q_{\mu\nu\rho} ~, 
\end{equation}
with
$\epsilon_{\mu\nu\rho\sigma} q^{\nu\rho\sigma} = q^\nu_{\,\rho\nu} =
0$.
This implies that the contorsion tensor undergoes the following decomposition:
\begin{equation}\label{hatted}
K_{abc} = \frac{1}{2} \epsilon_{abcd} S^d + {\widehat K}_{abc} 
\end{equation}
where $\widehat  K$ includes the  trace vector $T_\mu$ and  the tensor
$q_{\mu\nu\rho}$ parts of the torsion tensor.
 
The gravitational part of the action can then be written as:
\begin{equation}\label{toraction}
S_G =\frac{1}{2\kappa^2} \, \int d^4 x \sqrt{-g} \Big(R +
\widehat{\Delta} \Big) + \frac{3}{4\kappa^2} \int \textbf{S} \wedge
{}^\star\! \textbf{S}\; ,
\end{equation}
where  $\widehat \Delta  = {\widehat  K}^\lambda_{\ \: \mu\nu} {\widehat
  K}^{\nu\mu}_{\quad \lambda}  - {\widehat K}^{\mu\nu}_{\quad  \nu} \,
{\widehat K}^{\quad  \lambda}_{\mu\lambda}$, with the  hatted notation
defined in (\ref{hatted}).

In a  quantum gravity setting,  where one integrates over  all fields,
the torsion terms  appear as non propagating fields  and thus they can
be integrated out exactly. The authors of \cite{kaloper} have observed
though   that  the   classical  equations   of  motion   identify  the
axial-pseudovector torsion field $S_\mu$ with the axial current, since
the torsion equation yields
\begin{equation}\label{torsionec}
K_{\mu a b} = - \frac{1}{4} e^c_\mu \epsilon_{a b c d} \widebar{\psi}
\gamma_5 {\tilde \gamma}^d \psi\ .
\end{equation}
From this  it follows $\textbf{d}\,{}^\star\!\textbf{S}  = 0$, leading
to a  conserved ``torsion charge'' $Q =  \int {}^\star\!  \textbf{S}$.
To  maintain  this conservation  in  quantum  theory, they  postulated
$\textbf{d}\,{}^\star\!\textbf{S} = 0$ at the quantum level, which can
be  achieved  by  the  addition  of  judicious  counter  terms.   This
constraint, in a path-integral formulation of quantum gravity, is then
implemented  via a delta  function constraint,  $\delta (d\,{}^\star\!
\mathbf{S})$, and the latter via the well-known trick of introducing a
Lagrange multiplier  field $\Phi (x)  \equiv (3/\kappa^2)^{1/2} b(x)$.
Hence, the relevant torsion  part of the quantum-gravity path integral
would include a factor {\small
\begin{eqnarray}
 \label{qtorsion}
&&\hspace{-5mm}\int D \textbf{S} \, D b   \, \exp \Big[ i \int
    \frac{3}{4\kappa^2} \textbf{S} \wedge {}^\star\! \textbf{S} -
      \frac{3}{4} \textbf{S} \wedge {}^\star\! \textbf{J}^5  +
      \Big(\frac{3}{2\kappa^2}\Big)^{1/2} \, b \, d {}^\star\! \textbf{S}
      \Big]\nonumber \\  
&&\hspace{-5mm}=\!  \int D b  \, \exp\Big[ -i \int \frac{1}{2}
      \textbf{d} b\wedge {}^\star\! \textbf{d} b + \frac{1}{f_b}\textbf{d}b 
\wedge {}^\star\! \textbf{J}^5 + \frac{1}{2f_b^2}
    \textbf{J}^5\wedge\textbf{J}^5 \Big]\; ,\nonumber\\
\end{eqnarray}
\hspace{-1.5mm}}
where 
\begin{equation}\label{fbdef}
f_b = (3\kappa^2/8)^{-1/2} = \frac{M_P}{\sqrt{3\pi}}\  
\end{equation}
and  the  non-propagating   $\textbf{S}$  field  has  been  integrated
out. The reader  should notice that, as a  result of this integration,
the   corresponding   \emph{effective}   field   theory   contains   a
\emph{non-renormalizable} repulsive four-fermion axial current-current
interaction~\footnote{This  term  will  induce  a cubic  term  in  the
  equations  of motion  for  the fermions.   Under  the assumption  of
  formation of a (Lorentz violating) fermionic condensate of the axial
  current, it was recently argued~\cite{poplawski} that Dirac fermions
  may   lead  to   C-  and   CPT-violating  differences   between  the
  fermion-antifermion   populations    in   the   finite   temperature
  environment  of the  Early Universe.   In contrast,  for  a Majorana
  spinor of  interest to us, the  Majorana condition $\psi^c  = \psi $
  entails that a Majorana fermion is its own antiparticle and so there
  is \emph{no} such a violation of CPT.}.

We  may partially integrate  the second  term in  the exponent  on the
right-hand-side  of (\ref{qtorsion})  and take  into account  the well
known  field theoretic result  that in  QED the  axial current  is not
conserved at the  quantum level, due to anomalies,  but its divergence
is obtained by the one-loop result~\cite{anomalies}:
\begin{eqnarray}
   \label{anom}
\nabla_\mu J^{5\mu} \!&=&\! \frac{e^2}{8\pi^2} {F}^{\mu\nu}
  \widetilde{F}_{\mu\nu}  
- \frac{1}{192\pi^2} {R}^{\mu\nu\rho\sigma} \widetilde
{R}_{\mu\nu\rho\sigma} \nonumber\\ 
&\equiv& G(\textbf{A}, \omega)\; .
\end{eqnarray}
Observe that in (\ref{anom}) the torsion-free spin connection has been
used.  This can be achieved by the addition of proper counter terms in
the  action~\cite{kaloper}, which  can  convert the  anomaly from  the
initial    $G(\textbf{A},   \widebar   \omega)$    to   $G(\textbf{A},
\omega)$. Using  (\ref{anom}) in (\ref{qtorsion}) one  can then obtain
for the effective torsion action in QED
\begin{equation}\label{brr}
\int D b\ \exp\Big[ - i \int \frac{1}{2}
    \textbf{d} b\wedge {}^\star\! \textbf{d} b  - \frac{1}{f_b} b
    G(\textbf{A}, \omega)  
+ \frac{1}{2f_b^2} \textbf{J}^5 \wedge \textbf{J}^5 \Big]\; .
\end{equation}
Thus, even in  ordinary field theories, we obtain  the coupling of the
KR axion to the curvature  and gauge field strengths: $b G(\textbf{A},
\omega)$, exactly as we obtained in the string case (\ref{string}). In
addition,  the  torsion lead  to  repulsive four-fermion  interactions
involving  the axial  current. Crucial  to the  above  derivation was,
however, the postulation of the  conservation of the torsion charge at
the  quantum  level,  as   expressed  by  the  constraint  $\textbf{d}
{}^\star\!  \textbf{S} = 0$.  The resulting axion field has originated
from the  Lagrange multiplier field implementing  this constraint.  In
the subsequent  section we present  an alternative derivation  of this
result.

\subsection{Alternative Proof of the Connection of Axions to
  Torsion in Field Theory} 

We  shall now  provide a  different proof  concerning  the fundamental
geometrical properties  of the torsion  at the quantum level.  To this
end,  we concentrate on  the work  of \cite{mercuri},  which discusses
fermionic  torsion in  first-order Palatini  formalism of  fermions in
curved  space times.  Although  the  motivation of  that  work was  an
attempt to  connect some aspects  of loop quantum gravity  theories to
Ashtekar  canonical formulation of  quantum gravity,  nevertheless the
used  formalism   will  allow  us  to   discover  the  above-mentioned
torsion-induced axions from a different viewpoint.

As noted in~\cite{mercuri,Holst:1995pc}, in the case of Dirac fermions
in a manifold with torsion, if one uses the naive version of the Dirac
action (\ref{dirac}), decomposes the torsion tensor in its irreducible
parts, and  uses the equations of  motion, then the  trace vector part
$T_\mu$ is found proportional  to the axial fermion current $J^5_\mu$,
which   is    \emph{inconsistent}   with   the    respective   Lorentz
transformation properties. Such an inconsistency is remedied by adding
to the  action a\emph{ total  derivative} term which can  be expressed
solely in terms of \emph{topological invariants}, namely the so-called
Nieh-Yan invariant density~\cite{ny}, and a \emph{total divergence} of
the fermion axial current:
\begin{eqnarray}\label{ny}
S_{\rm Holst} &=& -i\frac{\eta}{2} \int d^4 x \left[I_{\rm NY} +
  \partial_\mu J^{5 \mu} \right]~, \nonumber \\ 
I_{\rm NY} & = & \epsilon^{\mu\nu\rho\sigma} \left[ K_{\,\,\mu\nu}^ a
  K_{\rho \sigma a } - \frac{1}{2} \Sigma_{\mu\nu}^{\,\,\ a b}\widebar{R}_{\rho
    \sigma a b } \right] \nonumber \\ & = &
\epsilon^{\mu\nu\rho\sigma} \partial_\mu K_{\nu \rho\sigma}~, 
\end{eqnarray}
where $\eta$  is a constant real  parameter, $\Sigma_{\mu\nu}^{\,\,\ a
  b} = \frac{1}{2}  e^a_{\left[\mu \right.} e^b_{\left. \nu \right]}$,
the  over line  above the  curvature tensor  denotes the  inclusion of
torsion  and   $K_{\mu\nu\rho}$  is  the   contorsion  tensor  defined
previously.   Notice  that  in  the  last equality  for  the  Nieh-Yan
topological  invariant we  took into  account the  fact that,  for the
torsion  (\ref{torsionec}), the  term quadratic  in the  contorsion in
$I_{\rm NY}$  vanishes\footnote{We note in passing  that the parameter
  $\eta$  in (\ref{ny})  is related  to the  so-called Immirzi-Barbero
  parameter   of   loop   quantum   gravity.   Such   completions   of
  torsionful-space-time gravity  theories, with the  above topological
  invariant terms, have been included for consistency in the canonical
  quantization of supergravity theories~\cite{kaul}, where the various
  fermion fields, including gravitinos, contribute to torsion.}. Then,
on  account of  (\ref{torsionec}), we  observe that  in the  case with
fermionic torsion, both terms in the space-time integrand of the Holst
action  turn out to  be proportional  to the  divergence of  the axial
current,  which  can  be  expressed  in terms  of  the  (torsion-free)
curvature  and  gauge field  strengths  through  the anomaly  equation
(\ref{anom})\footnote{In the case  of string-inspired H-torsion and in
  the presence of fermions,  the contortion tensor contains two parts,
  the   fermion-dependent   ones   (\ref{torsionec})  and   the   ones
  proportional to  the H-torsion.  In  that case, through  the anomaly
  equation  (\ref{anom}) and  Bianchi  identity (\ref{bianchi}),  both
  terms  in  the  expression  for  the  Holst  action  (\ref{ny})  are
  proportional  to  the  same  form  total  derivative  terms,  up  to
  algebraic  proportionality  factors.    The  final  result  for  the
  integrand of  the Holst  action is again  given by the  anomaly term
  (\ref{anom}),   $G(\mathbf{A},  \omega)$,   up   to  proportionality
  constants,  which  can  be  absorbed  in the  normalization  of  the
  parameter $\eta$.}.

By promoting the constant parameter $\eta$ into a pseudoscalar
field~\cite{mercuri2}, 
\begin{equation}
\eta\ \rightarrow\ \eta (x)\; ,  
\end{equation}
we notice that the term  involving the divergence of the axial current
in (\ref{ny}) yields, upon  using the anomaly equation (\ref{anom}), a
similar term  in the effective action  as the one  involving the axion
fields in string theory or in QED through implementing the appropriate
constraints by Lagrange multipliers, and  thus we can identify the non
constant field $\eta (x)$ with the axion $b(x)$
\begin{equation}
\eta (x)\ \equiv\ b(x)\; . 
\end{equation}
The  kinetic terms  for the  field $\eta$  are obtained  by  using the
equations  of  motion  and  identifying  extra  contributions  in  the
contorsion  involving derivatives of  the field.   This identification
has  also been  conjectured  in \cite{mercuri2}  without having  prior
knowledge of  the works  of \cite{kaloper}.  Indeed,  it was  shown in
\cite{mercuri2} that  the torsionful spin connection in  a theory with
fermions  and  a non  constant  $\eta  (x)  \equiv b(x)$  is  modified
compared to the constant $\eta$ case as follows:
\begin{equation}
\widebar  \omega_\mu^{\, \,a b}  =  \omega_\mu ^{\ \, a b}   (e)  +  \frac{1}{4}
\epsilon^{a b}_{\, \, \,\,\,cd} \, e_\mu^c \, \Big(\kappa  J^{5 d}  -  2 \eta^{d  f}
\partial_f b(x) \Big)\; ,
\end{equation}
where $\eta^{d f}$ is the  Minkowski metric on the tangent space.  The
quadratic   parts    of   the   torsionful    spin   connection   then
yield~\cite{mercuri2}  kinetic terms  for  the field  $b  (x)$ and  an
effective action  of the form (\ref{qtorsion}), which,  upon using the
anomaly equation  (\ref{anom}), implies the  axion-curvature couplings
mentioned above in (\ref{brr}).

Consequently, we  seem to  have established that  the presence  of the
coupling of axion  to spatial curvature and the  dual of the curvature
tensor is  a rather  generic feature of  torsionful theories  of space
time.  We  next proceed  to apply  the above ideas  to the  problem of
gauge invariant Majorana mass generation, without spontaneous symmetry
breaking, for right-handed Majorana  neutrinos, that carry no Standard
Model charges, and are thus susceptible to the effects of torsion.

\section{Anomalous Majorana Mass Generation from Quantum Torsion}

An important aspect of the coupling  of the torsion KR field $b(x)$ to
the  fermionic   matter  discussed   above  is  its   shift  symmetry,
characteristic of an axion field. Indeed, by shifting the field $b(x)$
by  a constant:  $b(x)  \to b(x)  +  c$, the  action (\ref{brr})  only
changes by  total derivative terms, such  as $c\, R^{\mu\nu\rho\sigma}
\widetilde{R}_{\mu\nu\rho\sigma}$               and               $c\,
F^{\mu\nu}\widetilde{F}_{\mu\nu}$.  These terms are irrelevant for the
equations  of motion and  the induced  quantum dynamics,  provided the
fields fall off sufficiently fast  to zero at space-time infinity. Our
scenario for  the anomalous  Majorana mass generation  through torsion
consists of augmenting the  effective action (\ref{brr}) by terms that
break such a shift symmetry.

To illustrate this last point, we  first couple the KR axion $b(x)$ to
another pseudoscalar  axion field $a(x)$.   In string-inspired models,
such  pseudoscalar   axion~$a(x)$  may  be  provided   by  the  string
moduli~\cite{arvanitaki,axiverse}.    The  proposed   coupling  occurs
through  a mixing  in  the kinetic  terms  of the  two  fields. To  be
specific, we consider the action
\begin{eqnarray} 
  \label{bacoupl}
    \mathcal{S} \!&=&\! \int d^4 x \sqrt{-g} \, \Big[\frac{1}{2}
      (\partial_\mu b)^2 + \frac{b(x)}{192 \pi^2 f_b}
      {R}^{\mu\nu\rho\sigma} \widetilde{R}_{\mu\nu\rho\sigma} \nonumber\\
      && + \frac{1}{2f_b^2} J^5_\mu {J^5}^\mu + \gamma
      (\partial_\mu b )\, (\partial^\mu a ) + \frac{1}{2}
      (\partial_\mu a)^2\nonumber\\ 
&&- y_a i a\, \Big( \widebar{\psi}_R^{\ C} \psi_R - \widebar{\psi}_R
\psi_R^{\ C}\Big) \Big]\;, \qquad 
\end{eqnarray}
where $\psi_R^{\ C} = (\psi_R)^C$ is the charge-conjugate right-handed
fermion $\psi_R$, $J_\mu^5  = \widebar{\psi} \gamma_\mu \gamma_5 \psi$
is the  axial current of  the four-component Majorana fermion  $\psi =
\psi_R  +  (\psi_R)^C$,  and  $\gamma$  is  a  real  parameter  to  be
constrained later on.   Here, we have ignored gauge  fields, which are
not of interest to us,  and the possibility of a non-perturbative mass
$M_a$  for the  pseudoscalar  field~$a(x)$.  Moreover,  we remind  the
reader that the {\em repulsive} self-interaction fermion terms are due
to the existence of torsion in the Einstein-Cartan theory.  The Yukawa
coupling $y_a$ of  the axion moduli field $a$  to right-handed sterile
neutrino  matter $\psi_R$  may  be due  to  non perturbative  effects.
These terms \emph{break} the shift symmetry: $a \to a + c$.

Before proceeding with the  evaluation of the anomalous Majorana mass,
it is convenient to diagonalize  the axion kinetic terms by redefining
the KR axion field as follows:
\begin{equation}
  \label{krredef}
b(x) \rightarrow {b^\prime}(x) \equiv b(x) + \gamma a(x) 
\end{equation}
This implies that the effective action (\ref{bacoupl}) now becomes:
\begin{eqnarray} 
  \label{bacoupl2}
&& \mathcal{S} =  \int d^4 x \sqrt{-g} \, \Big[\frac{1}{2}
      (\partial_\mu b^\prime )^2 + \frac{1}{2} \Big(1- \gamma^2
\Big) \, (\partial_\mu a)^2
  \nonumber \\ && + \frac{1}{2f_b^2} J^5_\mu {J^5}^\mu  +
\frac{b^\prime (x) - \gamma a(x)}{192 \pi^2 f_b}  
{R}^{\mu\nu\rho\sigma} \widetilde{R}_{\mu\nu\rho\sigma}  \nonumber\\ 
&&- y_a i a\, \Big( \widebar{\psi}_R^{\ C} \psi_R - \widebar{\psi}_R
\psi_R^{\ C}\Big) \Big]\;.\qquad
\end{eqnarray}
Thus we  observe that the $b^\prime  $ field has decoupled  and can be
integrated out  in the  path integral, leaving  behind an  axion field
$a(x)$  coupled   both  to  matter   fermions  and  to   the  operator
$R^{\mu\nu\rho\sigma}   {\widetilde   R}_{\mu\nu\rho\sigma}$,  thereby
playing now the  r\^ole of the torsion field.   We observe though that
the approach is only valid for
\begin{equation}
  \label{gammaregion}
|\gamma|\ <\ 1\; , 
\end{equation}
otherwise the  axion field would  appear as a  ghost, \emph{i.e.}~with
the  wrong  sign  of  its  kinetic  terms,  which  would  indicate  an
instability  of  the  model.  This  is the  only  restriction  of  the
parameter $\gamma$.
\begin{figure}[t]
 \centering
  \includegraphics[clip,width=0.40\textwidth,height=0.15\textheight]{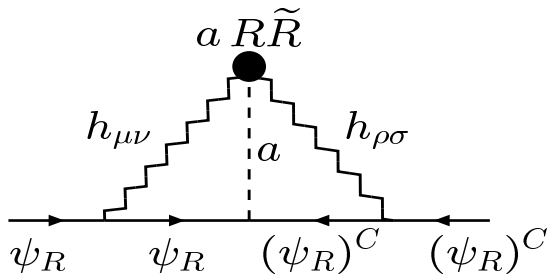} 
\caption{\it Typical Feynman graph giving rise to anomalous fermion
  mass generation.  The black circle denotes the operator $a(x)\,
  R_{\mu\nu\lambda\rho}\widetilde{R}^{\mu\nu\lambda\rho}$ induced by
  torsion.}\label{fig:feyn}
\end{figure}
In this case we may redefine the axion field so as to appear with a
canonical normalised kinetic term, implying the effective action:
{\small
\begin{eqnarray} 
  \label{bacoupl3} 
\mathcal{S}_a \!\!&=&\!\! \int d^4 x
    \sqrt{-g} \, \Big[\frac{1}{2} (\partial_\mu a )^2 - \frac{\gamma
        a(x)}{192 \pi^2 f_b \sqrt{1 - \gamma^2}}
      {R}^{\mu\nu\rho\sigma} \widetilde{R}_{\mu\nu\rho\sigma} \nonumber\\ 
&&\hspace{-5mm} - \frac{iy_a}{\sqrt{1 - \gamma^2}} \, 
a\, \Big( \widebar{\psi}_R^{\ C} \psi_R - \widebar{\psi}_R
\psi_R^{\ C}\Big) + \frac{1}{2f_b^2} J^5_\mu {J^5}^\mu
      \Big]\; .
\end{eqnarray}
\hspace{-1.5mm}}
Evidently, the  action $\mathcal{S}_a$ in~(\ref{bacoupl3}) corresponds
to a  canonically normalised axion field $a(x)$,  coupled \emph{both }
to the curvature of space-time, \emph{\`a la} torsion, with a modified
coupling $\gamma/(192 \pi^2  f_b \sqrt{1-\gamma^2})$, and to fermionic
matter  with  chirality-changing  Yukawa-like  couplings of  the  form
$y_a/\sqrt{1 - \gamma^2}$.

The mechanism for  the anomalous Majorana mass generation  is shown in
Fig.~\ref{fig:feyn}.   We  may  now  estimate  the  two-loop  Majorana
neutrino mass in  quantum gravity with an effective  UV energy cut-off
$\Lambda$.    Adopting  the   effective   field-theory  framework   of
\cite{Donoghue:1994dn},  we   first  notice  that   the  energy  ($E$)
dependence  of the  curvature  $R$  is $E^2$,  since  it contains  two
derivatives  $\partial_\mu$, with  $i\partial_\mu \to  p_\mu  \sim E$.
Therefore,              the              operator              $a(x)\,
R_{\mu\nu\lambda\rho}\widetilde{R}^{\mu\nu\lambda\rho}$  gives rise to
an  $E^4$ dependence.   Likewise,  on naive  dimensional grounds,  the
couplings    of   the    linearized    gravitons   $h_{\mu\nu}$    and
$h_{\rho\sigma}$ to  chiral fermions $\psi_R$ and  $\psi_R^{\ C}$ both
grow as $E$, as their kinetic  terms are proportional to a single power
of $i\partial_\mu$.   This gives rise to  another energy factor~$E^2$.
Collecting  all the  energy factors  resulting from  the gravitational
interactions and the loop momenta,  we find that the two-loop graph in
Fig.~\ref{fig:feyn}  exhibits  a  UV  cut-off  dependence~$\Lambda^6$.
This leads to a gravitationally induced Majorana mass~$M_R$:
\begin{equation}
  \label{MR}
M_R \sim \frac{1}{(16\pi^2)^2}\;
\frac{y_a\, \gamma\  \kappa^4 \Lambda^6}{192\pi^2 f_b (1 - \gamma^2 )} =
\frac{\sqrt{3}\, y_a\, \gamma\,  \kappa^5 \Lambda^6}{49152\sqrt{8}\,
\pi^4 (1 - \gamma^2 )}\; ,  
\end{equation}
where in the second step we  took into account (\ref{fbdef}). In a UV
complete theory  such as  strings, $\Lambda$  and $M_P$  are related,
since $\Lambda$ is proportional to  $M_s$ and the latter is related to
$M_P$ (or  $\kappa$) through (\ref{stringplanck}).

It is interesting  to provide a numerical estimate  of the anomalously
generated Majorana mass $M_R$. Assuming that $\gamma \ll 1$, the size 
of $M_R$ may be estimated from~(\ref{MR}) to be
{\small
\begin{equation}
  \label{MRest}
M_R \sim (3.1\times 10^{11}~{\rm GeV})\bigg(\frac{y_a}{10^{-3}}\bigg)\;
\bigg(\frac{\gamma}{10^{-1}}\bigg)  
\bigg(\frac{\Lambda}{2.4 \times 10^{18}~{\rm GeV}}\bigg)^6\, .
\end{equation}
\hspace{-1.5mm}}
Obviously, the generation of $M_R$ is highly model dependent.  Taking,
for example, the quantum gravity  scale to be $\Lambda = 10^{17}$~GeV,
we  find that  $M_R$ is  at the  TeV scale,  for $y_a  =  10^{-3}$ and
$\gamma =  0.1$. However, if we  take the quantum gravity  scale to be
close  to the  GUT scale,  i.e.~$\Lambda =  10^{16}$~GeV, we  obtain a
right-handed neutrino  mass $M_R \sim  16$~keV, for the choice  $y_a =
\gamma = 10^{-3}$.   This is in the preferred  ballpark region for the
sterile   neutrino    $\psi_R$   to    qualify   as   a    warm   dark
matter~\cite{Asaka:2006ek}.

In a  string-theoretic framework, many  axions might exist  that could
mix  with each  other.  Such  a  mixing can  give rise  to reduced  UV
sensitivity of  the two-loop  graph shown in  Fig.~\ref{fig:feyn}.  To
make this point explicit, let  us therefore consider a scenario with a
number $n$ axion fields, $a_{1,2,\dots,n}$.  Of this collection of $n$
pseudoscalars, only $a_1$ has a  kinetic mixing term $\gamma$ with the
KR  axion  $b$  and  only   $a_n$  has  a  Yukawa  coupling  $y_a$  to
right-handed neutrinos  $\psi_R$.  The other  axions $a_{2,3,\dots,n}$
have a next-to-neighbour mixing pattern.  In such a model, the kinetic
terms of the effective action are given by {\small
\begin{eqnarray}
  \label{Skin}
\mathcal{S}^{\rm  kin}_a \!\!&  =  &\!\! \int  d^4x\, \sqrt{-g}\,  \bigg[
  \frac{1}{2}\,\sum\limits_{i=1}^n \Big( (\partial_\mu   a_i)^2 - M^2_i\Big)  
+ \gamma  (\partial_\mu
  b)(\partial^\mu a_1)\nonumber\\
\!\!& &\!\! - \frac{1}{2}\,\sum\limits_{i=1}^{n-1} \delta M^2_{i,i+1}\,
a_i a_{i+1}\ \bigg]\; ,
\end{eqnarray}
\hspace{-1.5mm}}
where the mixing mass terms $\delta M^2_{i,i+1}$ are constrained to be
$\delta M^2_{i,i+1} < M_i M_{i+1}$,  so as to obtain a stable positive
mass   spectrum   for  all   axions.    As   a   consequence  of   the
next-to-neighbour mixing, the UV behaviour of the off-shell transition
$a_1 \to a_n$, described by the propagator matrix element $\Delta_{a_1
  a_n} (p)$,  changes drastically, i.e.~$\Delta_{a_1  a_n} (p) \propto
1/(p^2)^n \sim  1/E^{2n}$.  Assuming,  for simplicity, that  all axion
masses  and  mixings  are  equal,  i.e.~$M^2_i =  M^2_a$  and  $\delta
M^2_{i,i+1} =  \delta M^2_a$, the anomalously  generated Majorana mass
may be estimated to be
\begin{equation}
  \label{MRmix1}
M_R \sim 
\frac{\sqrt{3}\, y_a\, \gamma\,  \kappa^5 \Lambda^{6-2n} 
(\delta M^2_a)^n}{49152\sqrt{8}\, 
\pi^4 (1 - \gamma^2 )}\; ,
\end{equation}
for $n \leq 3$, and 
\begin{equation}
  \label{MRmix2}
M_R \sim 
\frac{\sqrt{3}\, y_a\, \gamma\,  \kappa^5 (\delta M^{2}_a)^3}{49152\sqrt{8}\, 
\pi^4 (1 - \gamma^2 )}\; \frac{(\delta
  M^{2}_a)^{n-3}}{(M^2_a)^{n-3}}\; ,
\end{equation}
for  $n >  3$.  It  is then  not difficult  to see  that  three axions
$a_{1,2,3}$ with next-to-neighbour mixing  as discussed above would be
sufficient  to obtain a  UV finite  result for  $M_R$ at  the two-loop
level. Of  course, beyond the two  loops, $M_R$ will  depend on higher
powers of the energy  cut-off $\Lambda$, i.e.~$\Lambda^{n> 6}$, but if
$\kappa\Lambda \ll  1$, these higher-order effects are  expected to be
subdominant.

In the above $n$-axion-mixing  scenarios, we note that the anomalously
generated  Majorana mass  term  will only  depend  on the  mass-mixing
parameters $\delta M_a^2$ of the  axion fields and not on their masses
themselves, as long as $n \le 3$.  Instead, for axion-mixing scenarios
with $n > 3$, the induced Majorana neutrino masses are proportional to
the factor $(\delta M^2_a/M^2_a)^n$, which gives rise to an additional
suppression for heavy axions with masses $M_a \gg \delta M_a$.

The possible existence of heavy pseudoscalar fields, with masses $M_a$
as   large  as   {\rm  TeV},   has  been   considered,   for  instance
in~\cite{heavy},   within   complicated   GUTs,  with   extra   strong
interactions,    confining   at    very    short   distance    scales.
Phenomenological implications and  astro\-physical constraints of such
heavy axions have been analyzed in~\cite{heavytests}, for a wide range
of  axion couplings.   Specifically, for  the Peccei-Quinn  (PQ) axion
scale~\footnote{The PQ  symmetry breaking scale  $f_a$ usually defines
  the strength of the axion interactions with the SM matter (nucleons)
  and  photons.  Adopting  the  notations  of  \cite{heavytests},  the
  relevant interaction Lagrangian may be written down as
\begin{displaymath}
{\mathcal  L}_{\rm  int}\  =\  -  i  y_{ak}  a(x)  \widebar{\psi}_k
  \gamma_5  \psi_k  +  \frac{1}{4} g_{a\gamma\gamma}  a(x)  F_{\mu\nu}
        \widetilde{F}^{\mu\nu}\:  +\: \dots\; ,
\end{displaymath}
where  $y_{ak} =  C_k m_k/f_a$  is the  Yukawa coupling  of  the axion
$a(x)$  to  the fermion  (nucleon)  species  $\psi_k$  of mass  $m_k$,
$g_{a\gamma\gamma}  = C_\gamma  \alpha_{\rm  em} /(2\pi  f_a)$ is  the
axion-photon-photon  coupling, and $C_{k,\gamma}$  are model-dependent
dimensionless parameters  which are  usually, but not  necessarily, of
order  one.  Notice  that there  is  no  bare  mass for  the  Majorana
fermion~$\psi$  in  our  case,  which the  pseudoscalar  field  $a(x)$
couples to.   Therofore, the  scale $f_a$ cannot  be defined,  but one
should  instead constrain directly  the dimensionless  Yukawa coupling
$y_a$  in  (\ref{bacoupl3}).  A  detailed  phenomenological study  may
appear elsewhere.   Of course,  one can always  assume that  the axion
moduli fields couple,  in addition to the Majorana  neutrinos, also to
the SM  matter, such as nucleons  and photons as above,  in which case
the  phenomenological analysis  and  constraints of  \cite{heavytests}
apply.}   $f_a  \sim   10^{12}$~GeV,  supernovae  data  exclude  heavy
pseudoscalar axion-like particles with  masses in the region $100~{\rm
  eV} \le  M_a \le 1~{\rm  GeV}$. Heavier axions are  allowed provided
their coupling is greater than $f_a \ge 10^{15}$~GeV.

In the  multi-axion models  outlined above, the  anomalously generated
Majorana neutrino mass $M_R$ will  still depend on the Yukawa coupling
$y_a$  and  the  torsion-axion  kinetic mixing  coefficient  $\gamma$,
besides the assumed UV  completion scale $\Lambda$ of quantum gravity.
In order to get  an estimate of the size of $M_R$,  we treat the axion
masses $M_a$  and mass-mixings $\delta  M_a$ as free parameters  to be
constrained by  phenomenology. Let us assume  a $n$-axion-mixing model
with $n\ge 3$, in which the  axion mass mixings $\delta M_a$ and their
masses $M_a$ are of the  same order, i.e.~$\delta M_a/M_a \sim 1$.  In
this  case,  employing~(\ref{MRmix1}), we  may  estimate the  Majorana
neutrino mass $M_R$ to be
\begin{eqnarray}
  \label{eq:multi}
\frac{M_R}{M_a} \!&\sim&\!  \frac{10^{-3}\,y_a\, \gamma}{1-\gamma^2}\,
\frac{(\delta  M_a)^6}{M_a \, M_P^5}\ \sim\ 
\frac{10^{-3}\, y_a\, \gamma}{1-\gamma^2} \bigg(\frac{\delta M_a}{M_P}\bigg)^5\ .\qquad
\end{eqnarray}
For axion  masses $M_a \le  1~{\rm TeV}$ considered in  the literature
thus far, we find that $M_R/M_a \stackrel{<}{{}_\sim} 10^{-83}$, which
implies extra-ordinary small Majorana masses $M_R$.  An obvious caveat
to this result  would be to have ultra-heavy  axion masses $M_a$ close
to  the GUT scale  and/or fine-tune  the torsion-axion  kinetic mixing
parameter   $\gamma$   to~1,  in   a   way   such   that  the   factor
$\frac{\gamma}{1-     \gamma^2}$    compensates    for     the    mass
suppression~$(\delta  M_a/M_P)^5$   in~(\ref{eq:multi}).   The  latter
possibility,   however,   might  result   in   an  unnaturally   large
(non-perturbative)  ``effective''   Yukawa  coupling  $y^{\rm  eff}_a\
\equiv\ y_a /\sqrt{1 - \gamma^2} \stackrel{>}{{}_\sim} \sqrt{4\pi}$ in
(\ref{bacoupl3}),  which  will   bring  us  outside  the  perturbative
framework  that  we  have  been  considering here.   A  more  detailed
phenomonological   and   astrophysical   analysis  of   all   possible
axion-mixing scenarios may be given elsewhere.

\section{Conclusions}

We have  shown how,  in theories of  quantum gravity with  torsion, an
effective right-handed  Majorana neutrino mass~$M_R$  can be generated
at  two  loops  by  gravitational  interactions  that  involve  global
anomalies.   The  global anomalies  result,  after  integrating out  a
formed-valued pseudoscalar field $b$, the so-called Kalb-Ramond axion,
which  describes the  effect of  quantum  torsion.  The  KR axion  $b$
couples to both matter and  to gravitation and radiation gauge fields.
In  perturbation  theory, this  torsion-descent  axion  field $b$  has
derivative couplings, leading  to an axion shift symmetry:  $b \to b +
c$, where $c$ is an arbitrary constant.  If another axion field $a$ or
fields are  present in the theory,  the shift symmetry  may be broken,
giving rise to axion masses and chirality changing Yukawa couplings to
massless fermions, such as right-handed Majorana neutrinos $\psi_R$.

We    have     estimated    the    magnitude     of    the    two-loop
gravitationally-induced  Majorana  mass $M_R$  and  found  that it  is
highly  model  dependent.   Its  size  generically  depends  on  three
parameters: the  value of the  Yukawa coupling $y_a$ to  $\psi_R$, the
kinetic mixing  term $\gamma$ between the  KR axion $b$  and the other
axion field $a$, and the assumed quantum gravity scale $\Lambda$ of UV
completion. In  the present study,  we have assumed that  $\Lambda$ is
considerably smaller  than the  Planck mass, i.e.~$\kappa  \Lambda \ll
1$. The  anomalously generated Majorana  neutrino mass $M_R$  can take
values, ranging from the multi-TeV to keV scale.

The radiative  fermion-mass mechanism discussed  in this paper  may be
used to  account for the  generation of other  gauge-invariant masses,
such as those pertinent to vector-like quarks and leptons. It would be
interesting to  investigate, whether  this radiative mechanism  can be
consistently extended to  super\-gravity theories with quantum torsion
and  analyze the possible  consequences of  such theories  on Majorana
fermion masses.

\bigskip

\noindent
{\bf Acknowledgements}\bigskip

The  work of  N.E.M. is  supported in  part by  the London  Centre for
Terauniverse Studies (LCTS), using  funding from the European Research
Council  via  the  Advanced  Investigator  Grant 267352,  and  by  the
U.K.~Science  and  Technology  Facilities  Council  (STFC)  under  the
research grant ST/J002798/1.   That of AP is supported  in part by the
Lancaster--Manchester--Sheffield  Consortium for  Fundamental Physics,
under STFC research grant: ST/J000418/1.

\vfill\eject


\begin{thebibliography}{99}

\bibitem{:2012gk} 
  G.~Aad {\it et al.}  [ATLAS Collaboration],
``Observation of a new particle in the search for the Standard Model
  Higgs boson with the ATLAS detector at the LHC,'' 
  Phys.\ Lett.\ B {\bf 716}, 1 (2012)
  [arXiv:1207.7214 [hep-ex]];\\
S.~Chatrchyan {\it et al.}  [CMS Collaboration],
``Observation of a new boson at a mass of 125 GeV with the CMS
  experiment at the LHC,'' 
  Phys.\ Lett.\ B {\bf 716}, 30 (2012)
  [arXiv:1207.7235 [hep-ex]].

\bibitem{seesaw}  P.~Minkowski, Phys.\ Lett.\ B {\bf 67}, 421 (1977);\\
M. Gell-Mann, P.  Ramond and R. Slansky, in  {\em Supergravity}, 
eds.~D.Z.  Freedman  and  P.~van Nieuwenhuizen  (North-Holland,  Amsterdam,
1979);\\ 
T.  Yanagida, in  Proc.\ of  the {\em  Workshop on  the Unified
Theory and the  Baryon Number in the Universe},  Tsukuba, Japan, 1979,
eds.\ O.~Sawada and  A.~Sugamoto;\\ 
R.~N.~Mohapatra and G.~Senjanovi\'c, Phys.\ Rev.\ Lett.\ {\bf 44}, 912 (1980).

\bibitem{Schechter:1980gr}
  J.~Schechter and J.~W.~F.~Valle,
  ``Neutrino Masses in SU(2) x U(1) Theories,''
  Phys.\ Rev.\ D {\bf 22}, 2227 (1980).

\bibitem{Mohapatra:2005wg}
  R.~N.~Mohapatra {\it et al.},
``Theory of Neutrinos: a White Paper,''
  Rept.\ Prog.\ Phys.\  {\bf 70}, 1757 (2007)
  [hep-ph/0510213].


\bibitem{Blumenhagen:2006xt} 
  R.~Blumenhagen, M.~Cvetic and T.~Weigand,
``Spacetime Instanton Corrections in 4D String Vacua: The Seesaw
  Mechanism for D-Brane Models,'' 
  Nucl.\ Phys.\ B {\bf 771}, 113 (2007)
  [hep-th/0609191];\\
  M.~Cvetic, R.~Richter and T.~Weigand,
``Computation of D-brane Instanton Induced Superpotential Couplings:
  Majorana Masses from String Theory,'' 
  Phys.\ Rev.\ D {\bf 76}, 086002 (2007)
  [hep-th/0703028].


\bibitem{ap} 
  A.~Pilaftsis,
  ``Anomalous Fermion Mass Generation at Three Loops,''
  arXiv:1207.0544 [hep-ph].
  
\bibitem{torsion}  F.~W.~Hehl, P.~Von Der Heyde, G.~D.~Kerlick and J.~M.~Nester,
  ``General Relativity with Spin and Torsion: Foundations and Prospects,''
  Rev.\ Mod.\ Phys.\  {\bf 48}, 393 (1976);\\
  I.~L.~Shapiro,
  ``Physical Aspects of the Space-Time Torsion,''
  Phys.\ Rept.\  {\bf 357}, 113 (2002)
  [hep-th/0103093] and references therein

\bibitem{kibble} T.~W.~B.~Kibble,
  ``Lorentz Invariance and the Gravitational Field,''
  J.\ Math.\ Phys.\  {\bf 2}, 212 (1961);
D.~W.~Sciama,
  ``The Physical Structure of General Relativity,''
  Rev.\ Mod.\ Phys.\  {\bf 36}, 463 (1964)
  [Erratum-ibid.\  {\bf 36}, 1103 (1964)].

\bibitem{tseytlin} 
  R.~R.~Metsaev and A.~A.~Tseytlin,
  ``Order alpha-prime (Two Loop) Equivalence of the String Equations
  of Motion and the Sigma Model Weyl Invariance Conditions: Dependence
  on the Dilaton and the Antisymmetric Tensor,'' 
  Nucl.\ Phys.\ B {\bf 293}, 385 (1987);\\ 
 D.~J.~Gross and J.~H.~Sloan,
  ``The Quartic Effective Action for the Heterotic String,''
  Nucl.\ Phys.\ B {\bf 291}, 41 (1987).

\bibitem{strominger} 
  A.~Strominger,
  ``Superstrings with Torsion,''
  Nucl.\ Phys.\ B {\bf 274}, 253 (1986).


\bibitem{kaloper}  M.~J.~Duncan, N.~Kaloper and K.~A.~Olive,
  ``Axion Hair and Dynamical Torsion from Anomalies,''
  Nucl.\ Phys.\ B {\bf 387}, 215 (1992).

\bibitem{Kalb:1974yc}
  M.~Kalb and P.~Ramond,
  ``Classical Direct Interstring Action,''
  Phys.\ Rev.\ D {\bf 9}, 2273 (1974).

\bibitem{poplawski} N.~J.~Poplawski,
  ``Matter--Antimatter Asymmetry and Dark Matter from Torsion,''
  Phys.\ Rev.\ D {\bf 83}, 084033 (2011)
  [arXiv:1101.4012 [gr-qc]].

\bibitem{anomalies} 
  R.~Delbourgo and A.~Salam,
``The Gravitational Correction to PCAC,''
  Phys.\ Lett.\ B {\bf 40}, 381 (1972);\\
see, also: S. Weinberg, \emph{The Quantum
  Theory of Fields. Volume II: Modern Applications.} (Cambridge
  University Press 2001) ISBN 0-521-55002-5. 


\bibitem{mercuri}   S.~Mercuri,
  ``From the Einstein-Cartan to the Ashtekar-Barbero Canonical
  Constraints, passing through the Nieh-Yan Functional,'' 
  Phys.\ Rev.\ D {\bf 77}, 024036 (2008)
  [arXiv:0708.0037 [gr-qc]] and references therein.

\bibitem{Holst:1995pc}
  S.~Holst, ``Barbero's Hamiltonian Derived from a Generalized Hilbert-Palatini Action,''
  Phys.\ Rev.\ D {\bf 53}, 5966 (1996) [gr-qc/9511026].

\bibitem{ny} H. T. ~Nieh and M.L.~Yan, ``An Identity in Riemann--Cartan
Geometry,'' J. Math.\ Phys.\ {\bf 23}, 373 (1982).


\bibitem{kaul} 
  R.~K.~Kaul,
  ``Holst Actions for Supergravity Theories,''
  Phys.\ Rev.\ D {\bf 77}, 045030 (2008)
  [arXiv:0711.4674 [gr-qc]];\\ 
  S.~Sengupta and R.~K.~Kaul,
  ``Canonical Supergravity with Barbero-Immirzi Parameter,''
  Phys.\ Rev.\ D {\bf 81}, 024024 (2010)
  [arXiv:0909.4850 [hep-th]].


\bibitem{mercuri2} V.~Taveras and N.~Yunes,
  ``The Barbero-Immirzi Parameter as a Scalar Field: K-Inflation from Loop
  Quantum Gravity?,''
  Phys.\ Rev.\  D {\bf 78}, 064070 (2008)
  [arXiv:0807.2652 [gr-qc]];\\
 G.~Calcagni and S.~Mercuri,
  ``The Barbero-Immirzi Field in Canonical Formalism of Pure Gravity,''
  Phys.\ Rev.\  D {\bf 79}, 084004 (2009)
  [arXiv:0902.0957 [gr-qc]];\\
S.~Mercuri and V.~Taveras,
  ``Interaction of the Barbero-Immirzi Field with Matter and
  Pseudo-Scalar Perturbations,'' 
  Phys.\ Rev.\ D {\bf 80}, 104007 (2009)
  [arXiv:0903.4407 [gr-qc]].
  
   
\bibitem{arvanitaki} 
  A.~Arvanitaki, S.~Dimopoulos, S.~Dubovsky, N.~Kaloper and J.~March-Russell,
  ``String Axiverse,''
  Phys.\ Rev.\ D {\bf 81}, 123530 (2010)
  [arXiv:0905.4720 [hep-th]].
    
 \bibitem{axiverse}  M.~Cicoli, M.~Goodsell and A.~Ringwald,
  ``The Type IIB String Axiverse and its Low-Energy Phenomenology,''
  arXiv:1206.0819 [hep-th].

\bibitem{Donoghue:1994dn}
  J.~F.~Donoghue,
  ``General Relativity as an Effective field theory: the Leading
  Quantum Corrections,'' 
  Phys.\ Rev.\ D {\bf 50}, 3874 (1994)
  [gr-qc/9405057].


\bibitem{Asaka:2006ek}
  T.~Asaka, M.~Shaposhnikov and A.~Kusenko,
  ``Opening a New Window for Warm Dark Matter,''
  Phys.\ Lett.\ B {\bf 638}, 401 (2006).
    
  \bibitem{heavy}   V.~A.~Rubakov,
  ``Grand Unification and Heavy Axion,''
  JETP Lett.\  {\bf 65}, 621 (1997)
  [hep-ph/9703409].
  
  
 \bibitem{heavytests} M.~Giannotti, L.~D.~Duffy and R.~Nita,
  ``New Constraints for Heavy Axion-like Particles from Supernovae,''
  JCAP {\bf 1101}, 015 (2011)
  [arXiv:1009.5714 [astro-ph.HE]];\\
  M.~Giannotti, R.~Nita and E.~Welch,
  ``Phenomenological Implications of Heavy Axion Models,''
  AIP Conf.\ Proc.\  {\bf 1274}, 20 (2010).
\end{thebibliography}
\end{document}